\documentclass[aps,prd,twocolumn,10pt]{revtex4-2}

\usepackage{amsmath,amssymb}
\usepackage{physics}
\usepackage{bbold}
\usepackage{t1enc}

\usepackage{amsfonts}
\usepackage{latexsym} 
\usepackage{cancel}
\usepackage{graphicx, rotating}
\usepackage{epsfig}
\usepackage{color}
\usepackage[dvipsnames]{xcolor}
\usepackage{multirow}

\DeclareMathSymbol{\shortminus}{\mathbin}{AMSa}{"39}
\DeclareMathSymbol{\shm}{\mathbin}{AMSa}{"39}

\begin{document}

\begin{center}
\begin{flushright}
IFT-UAM/CSIC-26-24
\end{flushright}
\end{center}
\vspace{-1cm}

\title{Quantum tomography of $H \to ZZ, WW$ beyond leading order}

\author{J. A.~Aguilar-Saavedra}
\affiliation{Instituto de F\'isica Te\'orica IFT-UAM/CSIC, c/Nicol\'as Cabrera 13--15, 28049 Madrid, Spain} 
\author{Pier Paolo Giardino}
\affiliation{Instituto de F\'isica Te\'orica IFT-UAM/CSIC, c/Nicol\'as Cabrera 13--15, 28049 Madrid, Spain} 
\affiliation{Departamento de Física Te\'orica, Universidad Aut\'onoma de Madrid, Cantoblanco, 28049, Madrid, Spain}

\begin{abstract}
We revisit quantum tomography of $H \to ZZ$ and $H \to WW$ in the presence of higher-order corrections. We verify that neither the use of an effective spin analysing power (only for $ZZ$) or a photon veto are sufficient to render the naively-constructed spin density operators physical. A subtraction of higher-order corrections is thus necessary to perform consistent quantum tomography. Such corrections are small when compared to expected experimental uncertainties with current data. As a by-product, we point out the striking possibility to observe parity-violating effects in $H \to WW$.
\end{abstract}

\maketitle

\section{Introduction}

Quantum tomography in high-energy collisions exploits the fact that, at leading order (LO), angular distributions of decay products encode the spin state of intermediate resonances. This correspondence enables the reconstruction of spin density operators~\cite{Kane:1991bg,Aguilar-Saavedra:2015yza,Aguilar-Saavedra:2017zkn,Rahaman:2021fcz,Ashby-Pickering:2022umy,Bernal:2023jba} and the study of quantum correlations such as entanglement~\cite{Afik:2020onf,Severi:2021cnj,Afik:2022kwm,Aoude:2022imd,Aguilar-Saavedra:2022uye,Afik:2022dgh,Severi:2022qjy,Dong:2023xiw,Han:2023fci,Maltoni:2024tul,Maltoni:2024csn,Cheng:2024btk,Aoude:2025ovu,Barr:2021zcp,Fabbrichesi:2023cev,Morales:2023gow,Aoude:2023hxv,Bernal:2023ruk,Fabbri:2023ncz,Bernal:2024xhm,Ruzi:2024cbt,Wu:2024ovc,Han:2024ugl,Bernal:2025zqq,Ding:2025mzj,Altakach:2022ywa,Ehataht:2023zzt,Fabbrichesi:2024wcd,Fabbrichesi:2024wcd,Han:2025ewp,Zhang:2025mmm,Afik:2025grr,Aguilar-Saavedra:2023hss,Aguilar-Saavedra:2024fig,Aguilar-Saavedra:2024hwd,Aguilar-Saavedra:2024vpd,Aguilar-Saavedra:2024whi,Aguilar-Saavedra:2023lwb,Liu:2025iwh,Afik:2026pxv}.
However, as pointed out in previous work~\cite{Aguilar-Saavedra:2025byk}, this framework is intrinsically limited to LO, since higher-order corrections introduce amplitudes that cannot be mapped onto well-defined spin degrees of freedom of the intermediate particles.

At next-to-leading order (NLO), virtual corrections and real radiation generically spoil the one-to-one correspondence between angular coefficients and spin operators. In concrete cases, such as 
$H \to ZZ^*$~\cite{Grossi:2024jae,DelGratta:2025qyp,Goncalves:2025mvl} and $H \to WW^*$~\cite{Goncalves:2025xer} this manifests itself in angular coefficients that, if interpreted naively, lead to non-physical spin density operators (not positive semidefinite) and invalidate any attempt to perform quantum tomography.\footnote{In these decays at least one boson is away from its mass shell because $M_H < 2 M_{Z,W}$. For brevity, we drop the superscript in the following, referring to the processes as $H \to ZZ$ or $H \to WW$.} A possible solution is to consider higher-order corrections, defined as the difference between LO and NLO differential cross sections,
\begin{equation}
\Delta_\text{NLO} \equiv d\sigma_\text{NLO} - d\sigma_\text{NLO}
\label{ec:D1}
\end{equation}
as a background and subtracted from data~\cite{Aguilar-Saavedra:2025byk}. 

Interestingly, Ref.~\cite{Goncalves:2025mvl} pointed out that a subset of NLO contributions to $H \to ZZ$ still admits a meaningful spin interpretation, and can be accounted for by a redefinition of the $Z$ spin-analysing power $\eta_\ell$. Then, it is reasonable to exclude them from the subtraction term $\Delta_\text{NLO}$ in (\ref{ec:D1}), thereby refining the subtraction approach of Ref.~\cite{Aguilar-Saavedra:2025byk}. This is not possible, however, for $H \to WW$, where $\eta_\ell = \pm 1$ at the tree level as well as at NLO.

In this work we reconsider both decays $H \to ZZ$ and $H \to WW$. We obtain the `naive' NLO density operators defined by applying quantum tomography to NLO angular distributions. We verify that neither using an effective $\eta_\ell$ (only for $ZZ$) nor imposing a veto on photon radiation (for both processes) render these naive operators positive semidefinite. 

An interesting byproduct of our analysis is the identification of parity-violation effects in $H \to WW$. For the charge-parity (CP) self-conjugate final states $ZZ$, $W^+ W^-$, parity (P) and CP are equivalent, and the small size of CP violation in Higgs decays, of the order of $10^{-5}$~\cite{Gritsan:2022php}, bars the appearance of observable P-violating effects. However, this is no longer the case in the presence of an extra photon.

In addition, we estimate the expected statistical precision for the determination of angular coefficients, which provides a lower bound on experimental uncertainties. This is an essential ingredient to assess the relevance of higher-order corrections. NLO effects for $H \to ZZ$ are of the order of the current experimental precision, and depending on systematic uncertainties, they may be also relevant for $H \to WW$.

When experimental precision requires higher-order calculations, physical density operators can be recovered by treating higher-order effects as a background~\cite{Aguilar-Saavedra:2025byk}. 
A detailed discussion of the meaning and conditions for higher-order subtraction is finally presented.

\section{$H \to V_1 V_2$ angular distributions}
\label{sec:2}

For $H \to ZZ$ we consider the $e^+ e^- \mu^+ \mu^-$ final state, so that the two intermediate $Z$ bosons can be {\em defined} from the two opposite-sign same-flavour pairs, with momenta $p_{Z_1} = p_{\ell_1^+} + p_{\ell_1^-}$, $p_{Z_2} = p_{\ell_2^+} + p_{\ell_2^-}$, $\ell = e,\mu$.\footnote{This assignment is not possible in the $4e/4\mu$ final state because of interference and identical-particle effects~\cite{Aguilar-Saavedra:2024jkj}.} For definiteness we label as $Z_1$ the one with largest invariant mass.
For $H \to W^+ W^-$ we consider the $e^\pm \nu \mu^\mp \nu$ final state to avoid contamination from $ZZ$-mediated amplitudes into same-flavour final states. In principle these would be suppressed by a kinematical reconstruction, however, the extent of this suppression in an experimental analysis is yet to be determined.

We parameterise angular distributions using the helicity basis, a moving reference system with vectors $(\hat r, \hat n, \hat k)$ defined as follows~\cite{Bernreuther:2015yna}:
\begin{itemize}
\item $\hat k$ is chosen in the direction of the $V_1 = Z_1,W^+$ momentum, evaluated either in the Higgs rest frame or the four-lepton c.m. frame. Both choices are equivalent at LO, but not in the presence of an extra photon. 
\item $\hat r$ is defined as $\hat r = \mathrm{sign}(\cos \theta) (\hat p_p - \cos \theta \hat k)/\sin \theta$, with $\hat p_p = (0,0,1)$ the direction of one proton in the laboratory frame, and $\cos \theta = \hat k \cdot \hat p_p$. The definition for $\hat r$  is the same if we use the direction of the other proton $- \hat p_p$. 
\item $\hat n$ is taken orthogonal, $\hat n = \hat k \times \hat r$.
\end{itemize}
The angular orientation of the decay products can be specified by the polar and azimuthal angles $\Omega_1 = (\theta_1,\phi_1)$ and $\Omega_2 = (\theta_2,\phi_2)$ of the charged lepton momenta from $V_1 V_2$ decay, measured in the $(\hat r, \hat n, \hat k)$ reference system. (For $V_1 V_2 = ZZ$ we use the negative leptons.) These leptons are often referred to as spin analysers. The four-dimensional decay angular distribution reads~\cite{Aguilar-Saavedra:2022wam,Aguilar-Saavedra:2022mpg}
\begin{eqnarray}
\frac{1}{\sigma}\frac{d\sigma}{d\Omega_1d\Omega_2} & = & \frac{1}{(4\pi)^2}\left[ 1 +a_{LM}^1 Y_L^M(\Omega_1) + a_{LM}^2  Y_L^M(\Omega_2)   \right. \notag \\
& & \left. + c_{L_1 M_1 L_2 M_2} Y_{L_1}^{M_1}(\Omega_1)Y_{L_2}^{M_2}(\Omega_2)  \right] \,,
\label{ec:dist4D}
\end{eqnarray}
with $Y_L^M$ the usual spherical harmonics, and implicit sum over repeated indices.
At LO, this expansion is general with $L \leq 2$, and the only non-zero coefficients are $a_{20}^{1,2}$, $c_{LML-M}$ with $M \leq L$. At higher orders, higher-rank spherical harmonics may in principle have significant contributions. For $H \to ZZ$, entries with $2 < L \leq 4$ are compatible with zero within Monte Carlo uncertainty; for $H \to WW$ there are a few entries at the percent level. Those contributions are also removed by the subtraction of the NLO corrections that is required to render the density operator physical.

For the calculation of $H \to ZZ \to e^+ e^- \mu^+ \mu^-$ and $H \to WW \to e^\pm \nu \mu^\mp \nu$ with electroweak corrections we use {\scshape MadGraph5\_aMC@NLO} with the same setup of Ref.~\cite{DelGratta:2025qyp}. We take as fundamental parameters for renormalisation
the masses of the weak bosons $M_Z = 91.188$ GeV, $M_W = 80.419$ GeV, and the Fermi constant $G_F = 1.16639 \times 10^{-5}$ GeV$^{-2}$, and use the complex-mass scheme~\cite{Denner:2006ic}.  The masses of the top quark and Higgs boson are taken  as $m_t = 173.3$ GeV, $M_H = 125$ GeV. Photon recombination is performed by clustering photons and leptons into `dressed' leptons if their angular separation $\Delta R \equiv [ (\Delta \phi)^2 + (\Delta \eta)^2]^{1/2}$ is smaller than 0.1. 
For $H \to ZZ$, it has been verified~\cite{DelGratta:2025qyp} that a smaller threshold results in larger differences between NLO and LO predictions for $a$ and $c$ coefficients, while a larger threshold largely recovers the LO values. The same behaviour has been observed for $H \to WW$~\cite{Goncalves:2025xer}. This feature is expected, as some important deviations from the LO pattern are caused by photon radiation off the final leptons. Therefore, by increasing $\Delta R$ the original lepton momenta before radiation are restored. On the other hand, defining dressed leptons with large $\Delta R$ may introduce significant contamination in a hadron collider environment. A veto on energetic photons has the same effect and experimentally seems more convenient.

It has been pointed out~\cite{Grossi:2024jae} that using a coordinate system where $\hat k$ is evaluated in the four-lepton c.m. frame, the differences between the LO and NLO angular coefficients are smaller than using the Higgs rest frame. For brevity, we present our results with the former choice. In addition, one can choose to veto events for which the photon is sufficiently energetic and observable. We will present results both inclusive (without veto) and exclusive (with an upper cut of 10 GeV on the photon energy). Previous literature has only provided results without veto~\cite{Grossi:2024jae,DelGratta:2025qyp,Goncalves:2025mvl,Goncalves:2025xer}, with the exception of $H \to ZZ$ in Ref.~\cite{Aguilar-Saavedra:2025byk}. Rejecting events with energetic photons is sensible because hard photon emission corresponds to a distinct physical process, and can be experimentally distinguished without significant cross section decrease: the efficiency of this cut, evaluated at the parton level, is 0.994 for $ZZ$ and 0.999 for $WW$.

We present in Table~\ref{tab:acZZ} the values of $a_{20}^{1,2}$ and $c_{LML-M}$ computed for $H \to ZZ \to e^+ e^- \mu^+ \mu^-$ at LO and NLO. The coefficients omitted are compatible with zero within Monte Carlo uncertainties, using $2 \times 10^7$ unweighted events at LO and $1.2 \times 10^8$ weighted events at NLO. For comparison, the last column shows the expected statistical uncertainties evaluated for the inclusive case (for the exclusive selection they are essentially the same, see section~\ref{sec:4}), for Run 2$+$3 data. The difference between LO and NLO is irrelevant for current measurements except possibly for $a_{20}^2$---and this depending on the actual experimental sensitivity because, as we have stressed, the statistical uncertainty given here is a lower bound to the experimental uncertainty. We postpone a more detailed discussion to section~\ref{sec:4}.

\begin{table}[htb]
\begin{center}
\begin{tabular}{lccccc}
& LO & \multicolumn{2}{c}{NLO } & $\Delta_\text{stat}$ \\
&  & Inclusive & Exclusive  \\
$a_{20}^1$ & $-0.664$ & $-0.631$ & $-0.658$ & $0.21$ \\
$a_{20}^2$ & $-0.664$ & $-0.606$ & $-0.636$ & $0.20$ \\
$c_{111-1}$ & $0.285$ & $0.046$ & $0.056$ & $0.62$ \\
$c_{1010}$ & $-0.178$ & $-0.005$ & $-0.010$ & $0.68$ \\
$c_{222-2}$ & $0.730$ & $0.720$ & $0.727$ & $0.65$ \\
$c_{212-1}$ & $-1.180$ & $-1.180$ & $-1.178$ & $0.51$ \\
$c_{2020}$ & $1.785$ & $1.748$ & $1.775$ & $0.72$
\end{tabular}
\end{center}
\caption{Numerical value of selected coefficients of the distribution (\ref{ec:dist4D}) for $H \to ZZ$ at LO and NLO. The last two columns display the expected statistical uncertainties at Run 2$+$3, as obtained in section~\ref{sec:4}.}
\label{tab:acZZ}
\end{table}

The non-vanishing coefficients for $H \to WW$ are shown in Table~\ref{tab:acWW}. For this final state, we use samples of $2 \times 10^7$ unweighted events at LO and $2.5 \times 10^8$ weighted events at NLO. The appearance of P-violating nonzero entries $a_{10}^{1,2}$, $c_{1M2-M}$, and $c_{2M1-M}$ at NLO deserves special attention. We assess that these entries are not an artifact of Monte Carlo calculation by dividing the sample into two sets and comparing the values for each sample, to verify that they are similar. (The same procedure is followed in $H \to ZZ$, finding they are compatible with zero.) The photon veto decreases the size of P-violating effects, as expected. In contrast to Ref.~\cite{Goncalves:2025xer}, we find $c_{1-110}$ and $c_{2-221}$ compatible with zero. 

The last column of Table~\ref{tab:acWW} displays, for reference, the expected statistical uncertainty for Run 2$+$3 data, evaluated for the inclusive case. Clearly, this is quite an underestimation of experimental uncertainties. The reconstruction of the $W^+$, $W^-$ rest frames with two missing neutrinos is difficult, if not impossible, and we expect that systematic uncertainties will be dominant. Still, this shows that in this idealised scenario the difference between LO and NLO is at the $1\sigma$ level at most, provided one considers a veto on energetic photons.

\begin{table}[t]
\begin{center}
\begin{tabular}{lcccccc}
& LO & \multicolumn{2}{c}{NLO } & \multicolumn{2}{c}{$\Delta_\text{stat}$} \\
&  & Inclusive & Exclusive &  \\
$a_{10}^1$ & 0 & $-0.030$ & $-0.013$ & $0.016$ \\
$a_{10}^2$ & 0 & $0.030$ & $0.013$ & $0.015$ \\
$a_{20}^1$ & $-0.646$ & $-0.615$ & $-0.630$ & $0.015$ \\
$a_{20}^2$ & $-0.646$ & $-0.613$ & $-0.628$ & $0.016$ \\
$c_{111-1}$ & $-5.954$ & $-5.934$ & $-5.955$ & $0.037$ \\
$c_{1010}$ & $3.735$ & $3.800$ & $3.772$ & $0.047$ \\
$c_{222-2}$ & $0.736$ & $0.746$ & $0.751$ & $0.046$ \\
$c_{212-1}$ & $-1.189$ & $-1.196$ & $-1.194$ & $0.039$ \\
$c_{2020}$ & $1.776$ & $1.741$ & $1.753$ & $0.055$ \\
$c_{112-1}$ & 0 & $-0.044$ & $-0.019$ & $0.044$ \\
$c_{211-1}$ & 0 & $0.048$ & $0.024$ & $0.044$  \\
$c_{1020}$ & 0 & $0.073$ & $0.032$ & $0.051$  \\
$c_{2010}$ & 0 & $-0.067$ & $-0.026$ & $0.055$ 
\end{tabular}
\end{center}
\caption{Numerical value of selected coefficients of the distribution (\ref{ec:dist4D}) for $H \to WW$ at LO and NLO. The last two columns display the expected statistical uncertainties at Run 2$+$3, as obtained in section~\ref{sec:4}.}
\label{tab:acWW}
\end{table}

As previously mentioned, at NLO higher-rank spherical harmonics contribute to the differential $H \to WW$ cross section at the percent level. For illustration, we list in Table~\ref{tab:acWWH} the most significant coefficients. We have verified that their values in the two event subsamples are similar, thus their appearance is not caused by statistical fluctuations. One can also observe that, as expected, setting an upper cut on the photon energy suppresses these contributions and brings the differential distribution closer to the LO one.

\begin{table}[htb]
\begin{center}
\begin{tabular}{lcc}
& \multicolumn{2}{c}{NLO }  \\
& Inclusive & Exclusive  \\
$c_{113-1}$ & $-0.023$ & $-0.011$ \\
$c_{1030}$ & $0.030$ & $0.012$ \\
$c_{213-1}$ & $-0.016$ & $-0.008$ \\
$c_{2030}$ & $0.020$ & $0.011$ \\
$c_{311-1}$ & $-0.020$ & $-0.008$ \\
$c_{312-1}$ & $0.014$ & $0.006$ \\
$c_{3010}$ & $0.026$ & $0.008$ \\
$c_{3020}$ & $-0.013$ & $-0.005$ \\
\end{tabular}
\end{center}
\caption{Numerical value of selected coefficients of spherical harmonics with $L > 2$ in the distribution (\ref{ec:dist4D}) for $H \to WW$ at NLO.}
\label{tab:acWWH}
\end{table}

We finally remark that, so far, the quantities $a_{LM}^{1,2}$ and $c_{L_1 M_1 L_2 M_2}$ are coefficients of the angular distribution when expanded in terms of spherical harmonics. Their spin interpretation is discussed in the next section.

\section{$H \to V_1 V_2$ density operators}
\label{sec:3}

The spin density operator $\rho_{S_1 S_2}$ for two massive spin-1 bosons can be written as an expansion in irreducible tensors $T^L_M$~\cite{Aguilar-Saavedra:2022wam},
\begin{eqnarray}
\rho_{S_1 S_2} & = & \frac{1}{9}\left[ \mathbb{1}_3 \otimes \mathbb{1}_3 + A^1_{LM} T^L_M \otimes \mathbb{1}_3 + A^2_{LM} \mathbb{1}_3\otimes T^L_M \right. \notag \\
& & \left.  + C_{L_1 M_1 L_2 M_2}\ T^{L_1}_{M_1} \otimes T^{L_2}_{M_2} \right] \,,
\label{ec:rho}
\end{eqnarray}
with constants $A_{LM}^{1,2}$, $C_{L_1 M_1 L_2 M_2}$. This operator has unit trace by construction, and its hermiticity requires that coefficients satisfy $A_{LM}^{1,2} = (-1)^M A_{L-M}^{1,2}$, $C_{L_1 M_2 L_2 M_2} = (-1)^{M_1 + M_2} (C_{L1 -M_1 L_2 -M_2})^*$. We normalise $T^L_M$ such that $\Tr\left[T^L_M\; \left(T^L_M\right)^{\dagger}\right] = 3 $, where $\left(T^L_M\right)^{\dagger}=(-1)^M \, T^L_M $. For $L=1$ we have $T^1_{\pm 1}=\mp\sqrt{3}/2 \,  (J_1 \pm i J_2)$ and $T^1_0=\sqrt{3/2} \, J_3$, 
where $J_i$ are the usual angular momentum operators.  For $L=2$ they are defined as
\begin{eqnarray}
T^2_{\pm 2} & = & \frac{2}{\sqrt{3}}\, (T_{\pm 1}^1)^2 \,, \nonumber \\
T^2_{\pm 1} & = & \sqrt{\frac{2}{3}} \left[T_{\pm 1}^1T_{0}^1 + T_{0}^1T_{\pm 1}^1\right] \,, \nonumber\\
T^2_0 & = & \frac{\sqrt{2}}{3} \left[T_1^1T_{-1}^1 + T_{-1}^1T_{1}^1+2(T_{0}^1)^2\right] \,.
\end{eqnarray} 
In $H \to V_1 V_2$ at the tree level one of the weak bosons is off-shell, therefore its degrees of freedom do not correspond to a spin-1 particle. However, when coupled to massless external fermions  the scalar component vanishes~\cite{Berge:2015jra} and one can effectively consider the virtual boson as a spin-1 particle. In decays $Z \to \ell^+ \ell^-$, $W \to \ell \nu$, with $\ell = e,\mu$, this is an excellent approximation. The correspondence between the coefficients of the density operator and the angular distribution is
\begin{align}
& a_{LM}^{1,2} = B_L A_{LM}^{1,2}  \,, \notag \\
& c_{L_1 M_1 L_2 M_2} = B_{L_1} B_{L_2} C_{L_1 M_1 L_2 M_2} \,,
\label{ec:acAC}
\end{align}
with, $B_{1,2}$ are constants, $B_1 = - \sqrt{2\pi} \eta_\ell$ and $B_2 = \sqrt{2\pi/5}$. For leptonic $Z$ decays, using as spin analyser the negative lepton,
\begin{equation}
\eta_\ell = \frac{1-4 s_W^2}{1-4 s_W^2 + 8 s_W^4} \,,
\end{equation}
with $s_W$ the sine of the weak mixing angle.
For $W^\pm$ decays $\eta_\ell = \mp 1$, using as spin analyser the charged lepton, positive or negative.

\subsection{$H \to ZZ$}

For this decay channel it has been suggested in Ref.~\cite{Goncalves:2025mvl} that using an effective mixing angle, instead of its on-shell value $s_W^2 = 1 - M_W^2/M_Z^2$, captures part of the NLO corrections, those that involve virtual corrections to $Z \to \ell^+ \ell^-$ and photon radiation in the decay. These corrections respect the tree-level description in terms of a scalar decay into two spin-1 bosons.
More recently, it has been explicitly shown~\cite{DelGratta:2025xjp} that, as expected, for the decay of an on-shell $Z$ boson the NLO distributions can be described by using the LO description with an effective $\eta_\ell$ evaluated at $M_Z$. The reason behind is the fact that for on-shell $Z$ the mixed corrections involving production and decay are not enhanced by the nearly on-shell $Z$ propagator, and therefore have little impact on the angular distribution.

Table \ref{tab:sW} collects the numerical values of the on-shell definition, and the effective $s_W^2$ at the $Z$ pole and at $\sim 30$ GeV, as well as the resulting values of $\eta_\ell$. Using these values, we determine the coefficients $A_{LM}^{1,2}$ and $C_{L_1 M_1 L_2 M_2}$ by naively using the tree-level correspondence (\ref{ec:acAC}). Results are presented in Table~\ref{tab:ACZZ}. For LO we use the nominal value of $\eta_\ell$. For NLO we use (a) the nominal value; (b) the effective values, evaluated at $M_Z$ and $\sim30$ GeV for $Z_1$ and $Z_2$, respectively.\footnote{For $Z_2$ the invariant mass distribution is broad, with average around 30 GeV. Reference~\cite{Goncalves:2025mvl} used the same value for both bosons.}

\begin{table}[h]
\begin{center}
\begin{tabular}{lcc}
& $s_W^2$ & $\eta_\ell$ \\
Nominal (on-shell)           & 0.22247 & 0.21932 \\
Effective ($M_Z$)  & 0.23229 & 0.14098 \\
Effective ($\sim30$ GeV) & 0.23279 & 0.13704
\end{tabular}
\end{center}
\caption{Values of the on-shell and effective $s_W^2$, and the lepton spin analysing power $\eta_\ell$ used in this work.}
\label{tab:sW}
\end{table}

\begin{table}[htb]
\begin{center}
\begin{tabular}{lcccccc}
& LO & \multicolumn{2}{c}{NLO nominal} & \multicolumn{2}{c}{NLO effective} \\
&  & Inclusive & Exclusive & Inclusive & Exclusive \\
$A_{20}^1$ & $-0.592$ & $-0.563$ & $-0.587$ & $-0.563$ & $-0.587$ \\
$A_{20}^2$ & $-0.592$ & $-0.540$ & $-0.568$ & $-0.540$ & $-0.568$ \\
$C_{111-1}$ & $0.942$ & $0.153$ & $0.185$ & $0.381$ & $0.462$ \\
$C_{1010}$ & $-0.587$ & $-0.017$ & $-0.034$ & $-0.043$ & $-0.085$ \\
$C_{222-2}$ & $0.581$ & $0.573$ & $0.578$ & $0.573$ & $0.578$ \\
$C_{212-1}$ & $-0.939$ & $-0.939$ & $-0.938$ & $-0.939$ & $-0.938$ \\
$C_{2020}$ & $1.420$ & $1.391$ & $1.413$ & $1.391$ & $1.413$
\end{tabular}
\end{center}
\caption{Numerical value of selected coefficients of the expansion (\ref{ec:rho}) for $H \to ZZ$.}
\label{tab:ACZZ}
\end{table}

Using the effective $s_W^2$ brings the NLO prediction for $C_{1M1-M}$ closer to LO, as already pointed out~\cite{Goncalves:2025mvl}. Here we also observe that the photon veto not only brings these coefficients closer to LO, but the remaining ones as well. 
It is interesting to express the density operators also in the basis of spin eigenstates
\begin{equation}
\{ |\!+\!+\rangle \,, |\!+\!0\rangle \, |\!+\!-\rangle \,,
|0+\rangle \,, |00\rangle \, |0-\rangle \,,
|\!-\!+\rangle \,, |\!-\!0\rangle \, |\!-\!-\rangle \} \,.
\label{ec:basis}
\end{equation}
A pruning procedure is implemented to remove spurious entries that are nonzero only because of Monte Carlo statistical fluctuations. 
For this, we divide the total sample in two parts (1) and (2), and evaluate the two $9 \times 9$ density matrices independently. We compute an upper cut $M$ defined as
\begin{equation}
M = \operatorname{max} \left\{ \operatorname{abs} \left((\rho_{S_1 S_2})_{ij}^{(1)} - (\rho_{S_1 S_2})_{ij}^{(2)} \right) \right\} \,,
\end{equation}
that is, the maximum absolute difference between the entries in the two subsamples. The final density operator is obtained with the total sample, discarding entries smaller than $M$.

At LO, the density operator obtained is
\begin{equation}
\rho_{S_1 S_2}^\text{LO} = \left(
\begin{array}{ccccccccc}
0 & 0 & 0 & 0 & 0 & 0 & 0 & 0 & 0 \\
0 & 0 & 0 & 0 & 0 & 0 & 0 & 0 & 0 \\
0 & 0 & 0.195 & 0 & -0.314 & 0 & 0.193 & 0 & 0 \\
0 & 0 & 0 & 0 & 0 & 0 & 0 & 0 & 0 \\
0 & 0 & -0.314 & 0 & 0.611 & 0 & -0.312 & 0 & 0 \\
0 & 0 & 0 & 0 & 0 & 0 & 0 & 0 & 0 \\
0 & 0 & 0.193 & 0 & -0.312 & 0 & 0.194 & 0 & 0 \\
0 & 0 & 0 & 0 & 0 & 0 & 0 & 0 & 0 \\
0 & 0 & 0 & 0 & 0 & 0 & 0 & 0 & 0
\end{array}
\right) \,,
\end{equation}
and is positive semidefinite. This sparse structure follows from relations between spin coefficients~\cite{Aguilar-Saavedra:2022wam,Aguilar-Saavedra:2022mpg},
\begin{eqnarray}
A_{20}^1 & = & A_{20}^2 \,, \notag \\
C_{111-1} & = & - C_{212-1} \,, \notag \\
C_{1010} & = & -1 - \frac{1}{\sqrt 2} A_{20}^1 \,, \notag \\
C_{2020} & = & 1 - \frac{1}{\sqrt 2} A_{20}^1 \,, \notag \\
C_{222-2} & = & \frac{1}{\sqrt 2}(A_{20}^1 + 1 ) \,.
\label{ec:rels}
\end{eqnarray}
At NLO, the naive density operator obtained doing LO quantum tomography is, for the `best' case (effective $s_W^2$, photon veto),
\begin{widetext}
\begin{equation}
\rho_{S_1 S_2}^\text{NLO} = \left(
\begin{array}{ccccccccc}
0.082 & 0     &     0 &     0 &     0 &     0 &     0 &     0 & 0 \\
0     & 0     &     0 & 0.076 &     0 &     0 &     0 &     0 & 0 \\
0     &     0 & 0.114 &     0 &-0.231 &     0 & 0.192 &     0 & 0 \\
0     & 0.076 &     0 & 0     &     0 &     0 &     0 &     0 & 0 \\
0     &     0 &-0.231 &     0 & 0.605 &     0 &-0.234 &     0 & 0 \\
0     &     0 &     0 &     0 &     0 & 0     &     0 & 0.082 & 0 \\
0     &     0 & 0.192 &     0 &-0.234 &     0 & 0.111 &     0 & 0 \\
0     &     0 &     0 &     0 &     0 & 0.082 &     0 & 0     & 0 \\
0     &     0 &     0 &     0 &     0 &     0 &     0 &     0 & 0.087
\end{array}
\right) \,.
\end{equation}
\end{widetext}
It is not positive definite, and its smallest eigenvalue is sizeable, $\lambda_\text{min} = -0.082$. With the other choices for NLO in Table~\ref{tab:ACZZ} the values of $\lambda_\text{min}$ are even larger in absolute value. This shows that naively matching spin observables to angular coefficients is not consistent in any of these cases. We can also observe the appearance of new nonzero entries, originating from the breaking of tree-level relations among spin coefficients.

\subsection{$H \to WW$}

As we have noted, in this case $\eta_\ell = \mp 1$ for $W^\pm$ decays at NLO as well as at LO. We collect in Table~\ref{tab:ACWW} the nonzero coefficients $A_{LM}^{1,2}$, $C_{L_1 M_1 L_2 M_2}$ defined by the tree-level relation (\ref{ec:acAC}). As it has been previously pointed out~\cite{Goncalves:2025xer}, the NLO and LO coefficients are much closer than in the $H \to ZZ$ case; here we also show that the photon veto brings them even closer. In particular, the photon veto suppresses the $P$-violating coefficients that are zero at LO.

We follow the same pruning procedure outlined in the previous subsection to obtain the density operators in the basis (\ref{ec:basis}) of spin eigenstates. The LO spin operator reads
\begin{equation}
\rho_{S_1 S_2}^\text{LO} = \left(
\begin{array}{ccccccccc}
0 & 0 & 0 & 0 & 0 & 0 & 0 & 0 & 0 \\
0 & 0 & 0 & 0 & 0 & 0 & 0 & 0 & 0 \\
0 & 0 & 0.197 & 0 & -0.314 & 0 & 0.195 & 0 & 0 \\
0 & 0 & 0 & 0 & 0 & 0 & 0 & 0 & 0 \\
0 & 0 & -0.314 & 0 & 0.605 & 0 & -0.315 & 0 & 0 \\
0 & 0 & 0 & 0 & 0 & 0 & 0 & 0 & 0 \\
0 & 0 & 0.195 & 0 & -0.315 & 0 & 0.198 & 0 & 0 \\
0 & 0 & 0 & 0 & 0 & 0 & 0 & 0 & 0 \\
0 & 0 & 0 & 0 & 0 & 0 & 0 & 0 & 0
\end{array}
\right) \,,
\end{equation}
and is positive semidefinite. The NLO naive density operator with photon veto is 
\begin{widetext}
\begin{equation}
\rho_{S_1 S_2}^\text{NLO} = \left(
\begin{array}{ccccccccc}
0     & 0     &     0 &     0 &     0 &     0 &     0 &     0 & 0 \\
0     &-0.003 &     0 & 0.003 &     0 &     0 &     0 &     0 & 0 \\
0     &     0 & 0.201 &     0 &-0.316 &     0 & 0.199 &     0 & 0 \\
0     & 0.003 &     0 &-0.002 &     0 &     0 &     0 &     0 & 0 \\
0     &     0 &-0.316 &     0 & 0.598 &     0 &-0.317 &     0 & 0 \\
0     &     0 &     0 &     0 &     0 & 0.003 &     0 &-0.002 & 0 \\
0     &     0 & 0.199 &     0 &-0.317 &     0 & 0.200 &     0 & 0 \\
0     &     0 &     0 &     0 &     0 &-0.002 &     0 & 0.003 & 0 \\
0     &     0 &     0 &     0 &     0 &     0 &     0 &     0 & 0
\end{array}
\right) \,.
\end{equation}
\end{widetext}
It is nonphysical, as its lowest eigenvalue is $\lambda_\text{min} = -0.0056$. (We have verified that the same eigenvalue is obtained consistently in the two subsamples, before and after pruning.) In the inclusive case the minimum eigenvalue is $\lambda_\text{min} = -0.0128$. Therefore, naively matching spin observables to angular coefficients is not consistent for $H \to WW$ either.

\begin{table}[htb]
\begin{center}
\begin{tabular}{lcccc}
& LO & \multicolumn{2}{c}{NLO} \\
&  & Inclusive & Exclusive \\
$A_{10}^1$ & 0 & $-0.012$ & $-0.005$ \\
$A_{10}^2$ & 0 & $-0.012$ & $-0.005$ \\
$A_{20}^1$ & $-0.577$ & $-0.549$ & $-0.562$ \\
$A_{20}^2$ & $-0.577$ & $-0.548$ & $-0.560$ \\
$C_{111-1}$ & $0.948$ & $0.944$ & $0.948$ \\
$C_{1010}$ & $-0.595$ & $-0.604$ & $-0.600$ \\
$C_{222-2}$ & $0.586$ & $0.594$ & $0.597$ \\
$C_{212-1}$ & $-0.946$ & $-0.951$ & $-0.950$ \\
$C_{2020}$ & $1.413$ & $1.385$ & $1.395$ \\
$C_{112-1}$ & 0 & $-0.016$ & $-0.007$ \\
$C_{211-1}$ & 0 & $-0.017$ & $-0.009$ \\
$C_{1020}$ & 0 & $0.026$ & $0.011$ \\
$C_{2010}$ & 0 & $0.024$ & $0.009$
\end{tabular}
\end{center}
\caption{Numerical value of selected coefficients of the expansion (\ref{ec:rho}) for $H \to WW$.}
\label{tab:ACWW}
\end{table}

\section{Statistical uncertainties}
\label{sec:4}

The differences between LO and NLO predictions for angular coefficients must be contextualised with respect to the experimental precision. A lower bound on experimental uncertainties is given by the statistical uncertainty, which we estimate in this section.

We work at the parton level, but injecting approximate efficiencies of 0.7 for lepton detection. This efficiency accounts for the minimum transverse momentum ($p_T$) thresholds required. We do not include any trigger requirement, which is not significant for $H \to ZZ$ (the presence of four leptons, some of them with significant $p_T$, is expected to fulfill one or many of the trigger conditions for one, two, or three leptons~\cite{trigger}) but important for $H \to WW$. We evaluate the uncertainties for the exclusive selection. The efficiencies for the photon veto are nearly one and do not affect the statistical uncertainties (for comparison, results in section~\ref{sec:2} are calculated without the veto.) In any case, we stress again that we are providing here a lower bound on the experimental uncertainty.

The statistical uncertainty on the angular coefficients is obtained, as previously in Ref.~\cite{Aguilar-Saavedra:2025byk}, by performing pseudo-experiments.\footnote{Our approach here is different from Ref.~\cite{Aguilar-Saavedra:2025byk}. Here we estimate the uncertainty in angular coefficients without any type of correction nor interpretation.} In each pseudo-experiment, a random subset of $n_S$ signal events is drawn from the total event set, with $n_S$ the expected number of signal events for the given cross section and luminosity. 
We consider two benchmarks: Run $2 + 3$ at 13 / 13.6 TeV with a luminosity $L = 350$ fb$^{-1}$, and HL-LHC at 14 TeV with $L = 3$ ab$^{-1}$. The cross sections for $gg \to H \to e^+ e^- \mu^+\mu^- (\gamma)$ and $gg \to H \to e^\pm \mu^\mp \nu \nu (\gamma)$ are~\cite{Cepeda:2019klc,LHCHiggsCrossSectionWorkingGroup:2016ypw}
\begin{align}
& 13~\text{TeV}: && \sigma_{ZZ} = 2.86~\text{fb} \,, && \sigma_{WW} = 245~\text{fb} \,, \notag \\
& 13.6~\text{TeV}: && \sigma_{ZZ} = 3.08~\text{fb} \,, && \sigma_{WW} = 263~\text{fb} \,, \notag \\
& 14~\text{TeV}: && \sigma_{ZZ} = 3.22~\text{fb} \,, && \sigma_{WW} = 275~\text{fb} \,.
\end{align}
The central values and statistical uncertainties obtained for $N=10000$ pseudo-experiments are collected in Table~\ref{tab:PEWW}. We note that besides different efficiency factors and the possible signal suppression by kinematical selection, experimental statistical uncertainties will generally be larger than the ones presented here, because of statistical fluctuations in the background that we do not consider here.

\begin{table}[htb]
\begin{center}
\begin{tabular}{lccc}
& Run $2 + 3$ & HL-LHC & NLO (exclusive) \\
$a_{20}^1$ & $-0.67 \pm 0.20$ & $-0.658 \pm 0.068$ & -0.658
\\
$a_{20}^2$ &$-0.64 \pm 0.21$ & $-0.637 \pm 0.068$ & -0.636 
\\
$c_{111-1}$ &$0.05 \pm 0.64$ & $0.05 \pm 0.20$ & 0.056 
\\
$c_{1010}$ &$-0.02 \pm 0.70$ & $-0.03 \pm 0.23$ & -0.010 
\\
$c_{222-2}$ &$0.73 \pm 0.65$ & $0.73 \pm 0.21$ & 0.727
\\
$c_{212-1}$ &$-1.16 \pm 0.53$ & $-1.18 \pm 0.17$ & -1.178 
\\
$c_{2020}$ & $1.75 \pm 0.73$ & $1.78 \pm 0.23$ & 1.775 
\end{tabular}
\end{center}
\begin{center}
\begin{tabular}{lccc}
& Run $2 + 3$ & HL-LHC & NLO (exclusive) \\
$a_{10}^1$ & $-0.011 \pm 0.015$ & $-0.011 \pm 0.005$ & $-0.013$ \\
$a_{10}^2$ & $0.015 \pm 0.015$ & $0.014 \pm 0.005$ & $0.013$ \\
$a_{20}^1$ & $-0.632 \pm 0.016$ & $-0.631 \pm 0.005$ & $-0.630$ \\
$a_{20}^2$ & $-0.629 \pm 0.016$ & $0.629 \pm 0.005$ & $-0.628$ \\
$c_{111-1}$ & $-5.955 \pm 0.038$ & $-5.953 \pm 0.012$ & $-5.955$ \\
$c_{1010}$ & $3.758 \pm 0.050$ & $3.761 \pm 0.016$ & $3.772$ \\
$c_{222-2}$ & $0.759 \pm 0.048$ & $0.757 \pm 0.015$ & $0.751$ \\
$c_{212-1}$ & $-1.188 \pm 0.040$ & $-1.189 \pm 0.013$ & $-1.195$ \\
$c_{2020}$ & $1.739 \pm 0.056$ & $1.740 \pm 0.017$ & $1.753$ \\
$c_{112-1}$ & $-0.027 \pm 0.042$ & $-0.026 \pm 0.013$ & $-0.019$ \\
$c_{211-1}$ & $0.016 \pm 0.044$ & $0.016 \pm 0.014$ & $0.024$ \\
$c_{1020}$ & $0.034 \pm 0.052$ & $0.033 \pm 0.017$ & $0.032$ \\
$c_{2010}$ & $-0.028 \pm 0.053$ & $-0.027 \pm 0.017$ & $-0.026$
\end{tabular}
\caption{Central value and statistical uncertainty obtained for the angular coefficients in $H \to ZZ$ (top) and $H \to WW$ (bottom) from pseudo-experiments, in the exclusive selection. For comparison, the last columns show the NLO value.}
\label{tab:PEWW}
\end{center}
\end{table}

From these results, and having in mind that these estimations of statistical uncertainties are optimistic, we can conclude that NLO corrections are hardly necessary to describe data with current statistics. For the HL-LHC they will be relevant for $H \to ZZ$; however, for $H \to WW$ no conclusion can be drawn from this study, since systematic uncertainties are expected to dominate.

In the previous section we have noted that the density operator in the spin eigenstate basis (\ref{ec:basis}) has some entries that vanish at the tree level because of the relations (\ref{ec:rels}) but are nonzero at NLO. It is then relevant to examine the expected statistical uncertainty in the elements of the $9 \times 9$ matrices. With Run 2$+$3 statistics, we find
\begin{widetext}
\begin{equation}
\delta \rho_{S_1 S_2}^{ZZ} = \left(
\begin{array}{ccccccccc}
0.42 & 0.34 & 0.21 & 0.33 & 0.42 & 0.26 & 0.21 & 0.26 & 0.17 \\
0.34 & 0.25 & 0.32 & 0.42 & 0.20 & 0.41 & 0.27 & 0.13 & 0.26 \\
0.21 & 0.32 & 0.41 & 0.26 & 0.42 & 0.34 & 0.17 & 0.26 & 0.22 \\
0.33 & 0.42 & 0.26 & 0.24 & 0.20 & 0.13 & 0.33 & 0.41 & 0.26 \\
0.42 & 0.20 & 0.41 & 0.20 & 0.14 & 0.20 & 0.41 & 0.21 & 0.41 \\
0.26 & 0.41 & 0.34 & 0.13 & 0.20 & 0.24 & 0.26 & 0.40 & 0.34 \\
0.21 & 0.27 & 0.17 & 0.32 & 0.41 & 0.26 & 0.41 & 0.34 & 0.21 \\
0.26 & 0.13 & 0.26 & 0.41 & 0.21 & 0.40 & 0.34 & 0.24 & 0.33 \\
0.17 & 0.26 & 0.22 & 0.26 & 0.41 & 0.34 & 0.21 & 0.33 & 0.41
\end{array}
\right) \,,
\end{equation}
\begin{equation}
\delta \rho_{S_1 S_2}^{WW} = \left(
\begin{array}{ccccccccc}
0.003 & 0.004 & 0.006 & 0.004 & 0.006 & 0.009 & 0.005 & 0.009 & 0.012 \\
0.004 & 0.007 & 0.005 & 0.006 & 0.007 & 0.007 & 0.009 & 0.010 & 0.009 \\
0.006 & 0.005 & 0.005 & 0.009 & 0.007 & 0.005 & 0.012 & 0.009 & 0.006 \\
0.004 & 0.006 & 0.009 & 0.006 & 0.007 & 0.010 & 0.005 & 0.007 & 0.009 \\
0.006 & 0.007 & 0.007 & 0.007 & 0.011 & 0.007 & 0.007 & 0.007 & 0.006 \\
0.009 & 0.007 & 0.005 & 0.010 & 0.007 & 0.007 & 0.009 & 0.006 & 0.004 \\
0.005 & 0.009 & 0.012 & 0.005 & 0.007 & 0.009 & 0.005 & 0.005 & 0.006 \\
0.009 & 0.010 & 0.009 & 0.007 & 0.007 & 0.006 & 0.005 & 0.006 & 0.004 \\
0.012 & 0.009 & 0.006 & 0.009 & 0.006 & 0.004 & 0.006 & 0.004 & 0.003 \\
\end{array}
\right)
\end{equation} \,.
\end{widetext}
Therefore, it is safe to state that with current data, the appearance of nonzero entries in the $9 \times 9$ matrix at NLO is statistically insignificant, especially for $H \to ZZ$.

\section{Subtraction of higher orders}
\label{sec:5}

When higher-order effects invalidate a description in terms of intermediate resonances, a subtraction of these higher-order corrections is necessary to perform quantum tomography.
This approach is formally correct whatever the size of these corrections is. Instead of comparing experiment and theory
\begin{equation}
d\sigma_\text{exp} \leftrightarrow d\sigma_\text{th} \,,
\label{ec:comp}
\end{equation}
one can always subtract some gauge-invariant quantity on both sides,
\begin{equation}
d\sigma_\text{exp} - \Delta \leftrightarrow d\sigma_\text{th} - \Delta \,.
\label{ec:comp2}
\end{equation}
This approach has previously been used in the definition of $Z$-pole pseudo-observables $R_b$, $A_\text{FB}^b$, etc.~\cite{ParticleDataGroup:2024cfk}. These quantities are defined without the photon contributions, which are thereby subtracted when reporting experimental measurements~\cite{ALEPH:2005ab}.
In the same vein, we define suitable pseudo-observables where a gauge-invariant subset of higher order corrections is subtracted. In $H \to ZZ$ and $H \to WW$ higher-order corrections are small, $\Delta \ll d\sigma_\text{th}$, and the data-theory comparison in (\ref{ec:comp2}) keeps the same sensitivity to non-standard effects---beyond the SM or beyond quantum mechanics---as that in (\ref{ec:comp}).

A word of caution is necessary here. The statement that higher-order corrections are overall small does not preclude that their contribution to some conveniently crafted observable, like $c_{1010}$ and $c_{111-1}$, be large. An analogous situation is well known in the Standard Model (SM). It is often stated that CP-violating effects, which arise beyond the LO, are small. Still, for a CP-violating observable, the relative correction from the CP-conserving tree-level value (zero), to a CP-violating non-zero value at higher orders, is huge.

In principle, the $\Delta$ term in (\ref{ec:comp2}) includes corrections to all orders. When computing it at NLO, we are implicitly assuming that data is well described at NLO, and next-to-next-to-leading order (NNLO) calculations are not necesary. Otherwise, they should be included as well in $\Delta$. At NLO, $\Delta$ comprises the interference with LO as well as NLO corrections, 
\begin{eqnarray}
\Delta_\text{NLO} & \equiv & d\sigma_\text{NLO} - d\sigma_\text{LO} \notag \\
& = & d\Phi_F \; 2 \Re \mathcal{M}_F^{(1)} \mathcal{M}_F^{(0)*} \notag \\
& & + 
\int d\Phi_{F+1} \;  |\mathcal{M}_{F+1}^{(0)}|^2 \,,
\label{ec:DNLO}
\end{eqnarray}
where $\mathcal{M}_F^{(0)}$ and $\mathcal{M}_F^{(1)}$ are the amplitudes at tree-level and one loop, and $\mathcal{M}_{F+1}^{(0)}$ is the tree-level amplitude with an extra photon; $\Phi_F$ is the four-lepton phase space, and $\Phi_{F+1}$ the phase space of the four leptons and the photon, which we integrate over the photon degrees of freedom. From its definition, c.f. the first line of (\ref{ec:DNLO}), $\Delta_\text{NLO}$ is obviously gauge invariant.

The correction to angular coefficients stemming from (\ref{ec:DNLO}) is simply computed by subtracting the NLO and LO values---for the former, one needs to select a radius for the definition for dressed leptons and choose whether to veto on visible photons or not. For $H \to WW$, the correction can be directly read from Table~\ref{tab:acWW}. For $H \to ZZ$, this strategy can be further refined by incorporating a subset of higher-order corrections that preserve the two-qutrit description through the introduction of an effective $\eta_\ell$, as discussed in~\cite{Goncalves:2025mvl,DelGratta:2025xjp}. Consequently, the subtraction proposed in Ref.~\cite{Aguilar-Saavedra:2025byk} can be improved by evaluating the LO contribution in Eq.~(\ref{ec:DNLO}) using this effective $\eta_\ell$. This modification reduces the size of the NLO correction that must be subtracted to construct the pseudo-observable and has the potential to decrease the relative theoretical uncertainties.

The values of the angular coefficients at LO using the effective values of $s_W^2$ and $\eta_\ell$ in Table~\ref{tab:sW} are collected in Table~\ref{tab:acZZ2}. For better comparison, we also include the LO and NLO values from Table~\ref{tab:acZZ}, all these computed using the nominal value of $s_W^2$. As said, for the $\eta_\ell$-sensitive coefficients the difference between LO and NLO decreases when using at LO the effective values of $\eta_\ell$. 

\begin{table}[htb]
\begin{center}
\begin{tabular}{lccccc}
& \multicolumn{2}{c}{LO} & \multicolumn{2}{c}{NLO} \\
& Nominal & Effective & Inclusive & Exclusive  \\
$a_{20}^1$ & $-0.664$ & $-0.663$ & $-0.631$ & $-0.658$  \\
$a_{20}^2$ & $-0.664$ & $-0.662$ & $-0.606$ & $-0.636$  \\
$c_{111-1}$ & $0.285$ & $0.117$  & $0.046$ & $0.056$  \\
$c_{1010}$ & $-0.178$ & $-0.082$ & $-0.005$ & $-0.010$  \\
$c_{222-2}$ & $0.730$ & $0.729$  & $0.720$ & $0.727$ \\
$c_{212-1}$ & $-1.180$ & $-1.186$ & $-1.180$ & $-1.178$ \\
$c_{2020}$ & $1.785$ & $1.773$ & $1.748$ & $1.775$ 
\end{tabular}
\end{center}
\caption{Numerical value of selected coefficients of the distribution (\ref{ec:dist4D}) for $H \to ZZ$ at LO (using the nominal and effective $\eta_\ell$) and NLO with nominal $\eta_\ell$.}
\label{tab:acZZ2}
\end{table}

\section{Discussion}
When quantum corrections beyond LO are comparable in size to the experimental uncertainty, a dedicated scheme is required to perform quantum tomography that consistently incorporates these corrections. We have verified that for $H \to ZZ$ and $H \to WW$ neither the use of an effective $\eta_\ell$ nor a photon veto renders the naively constructed density operators physical. 

Alternatively, one may be tempted to sidestep the problem by defining a more general framework without resorting to intermediate resonances. For example, in $H \to V_1 V_2$ one could attempt to perform quantum tomography of the two dilepton pairs, irrespectively of whether they arise from intermediate resonances or not. For an $e^+ e^- \mu^+ \mu^-$ final state one would define $V_1$ as the $e^+ e^-$ pair and $V_2$ as the $\mu^+ \mu^-$ pair, and attempt to determine the spin properties of $V_1$ and $V_2$. This approach, however, is not viable. First, a lepton pair does not generally couple into a state of total angular momentum $J=1$; depending on the orbital angular momentum, configurations with $J=0,1,2,\dots$ are possible. Second, the relation between angular observables and spin observables, which stems from the chirality of the coupling and is well known at LO for $W$ and $Z$ bosons~\cite{Aguilar-Saavedra:2015yza,Aguilar-Saavedra:2017zkn}, cannot be straightforwardly computed for a lepton pair produced through arbitrary Feynman diagrams.

We therefore find it necessary to define new pseudo-observables in which the contributions from higher orders are subtracted. This relatively straightforward procedure makes it possible to recover the angular structure arising from the LO amplitude, thereby enabling a consistent implementation of quantum tomography. Further refinements, such as the introduction of an effective $\eta_\ell$, may then be employed to reduce the theoretical uncertainties associated with the subtraction procedure.

Finally, we have pointed out the remarkable possibility of observing P violation in Higgs decays. In particular, for $H \to WW$ we find the appearance of P-violating contributions to the differential cross section, corresponding to non-zero coefficients $a_{10}^{1,2}$, $c_{1M2-M}$, and $c_{2M1-M}$. The results presented indicate that the available statistics could allow for the measurement of these P-violating angular coefficients. However, these measurements—like the rest of the angular coefficients in $H \to WW$—are extremely challenging because the final state cannot be uniquely reconstructed due to the presence of two missing neutrinos, and systematic effects will certainly reduce the sensitivity. Nevertheless, this possibility deserves further investigation.

\section*{Acknowledgements}

This work has been supported by the Spanish Research Agency (Agencia Estatal de Investigaci\'on) through projects PID2022-142545NB-C21,  and CEX2020-001007-S funded by MCIN/AEI/10.13039/501100011033. P.P.G. is supported by the Ram\'on y Cajal grant RYC2022-038517-I funded by MCIN/AEI/10.13039/501100011033 and by FSE+.

\end{document}